\documentclass[a4paper, 12pt, doc]{apa7}

\usepackage[utf8]{inputenc}
\usepackage[english]{babel}
\usepackage{geometry}
\usepackage{csquotes}
\usepackage{orcidlink}
\usepackage{booktabs}
\usepackage{hyperref}
\usepackage{enumitem}
\usepackage{multirow}
\usepackage{graphicx}
\usepackage{float}
\usepackage{endnotes}
\let\footnote=\endnote
 
\usepackage[style=apa,backend=biber]{biblatex}
\title{Sustaining Dryad: Reconfiguring relationships and ‘thinking like a business’ to maintain open data infrastructures}
\shorttitle{Sustaining Dryad}

\authorsnames{Kathleen Gregory$^1$ \\ \orcidlinkf{0000-0001-5475-8632}
\\
,
\\
Dorothea Strecker$^2$ \\ \orcidlinkf{0000-0002-9754-3807}}

\authorsaffiliations{{$^1$Leiden University, Centre for Science and Technology Studies (CWTS)}, {$^2$Humboldt-Universität zu Berlin, Berlin School of Library and Information Science}}

\abstract{Open data infrastructures underpin how open research is practiced across communities. While critical, the longevity of such infrastructures is far from certain, as they grapple with challenges related to technologies, dominant ideologies, and continued funding streams. This paper draws on a mixed-methods study of Dryad, a well-known open data infrastructure, to deeply explore one of these challenges: the quest for stable funding. Our analysis shows how Dryad has carefully reconfigured different forms of relationships and revenue models throughout its history to work towards financial sustainability. We identify four types of relationships with customers, collaborators, and competitors that have been critical to Dryad’s financial evolution: \emph{reinforcing}, \emph{forging}, \emph{positioning}, and \emph{excluding} relationships. We argue that while implementing strategic, business thinking is a critical strategy, it is also one which shapes other factors important in sustainability: interpretations of value(s), community, and governance. We conclude by highlighting emerging tensions that provide insight for other open data infrastructures working to become financially sustainable. As a whole, our analysis focuses not just on financial mechanisms for funding open data infrastructures (although those emerge) but on the relationships which enable them.
\\
\\
\textbf{Keywords}: research data, open science, open infrastructures, data repositories, funding models, sustainability}

\begin{document}
\maketitle

\setlength{\parindent}{0pt}

\pagebreak

\section*{Media Summary}
Open data infrastructures (ODIs), such as research data repositories and federal governmental portals, underpin how data is made available. While critical, the longevity of such infrastructures is far from certain, as many continuously struggle to obtain stable funding. At the same time, the financial models for these infrastructures often remain hidden from view. Understanding more about these financial mechanisms – including their business models and the relationships which constitute them – is therefore critical to ensure the longevity of open data infrastructures. 

In this article, we present an in-depth study of the financial evolution of Dryad, an established open data infrastructure developed to facilitate data sharing and reuse. We combine semi-structured interviews, document review, and a quantitative analysis of Dryad’s collection to show how Dryad has worked toward financial sustainability by 1) reconfiguring relationships of different types with different actors and 2) explicitly engaging in ‘business thinking’ by identifying a value proposition, re-thinking costs and revenue sources, and turning curation services into an asset. 

While we detail the case of Dryad, our findings are significant for other open data infrastructures. In particular, our analysis highlights three important considerations. First, we find that while ‘thinking like a business’ is critical, it will also change other core aspects of an infrastructure related to value and values, modes of governance, and notions of community. Second, we argue that developing and maintaining strategic relationships with different actors is just as important as obtaining multiple sources of funding and that business models themselves can provide a means for managing such relationships. Finally, our analysis raises questions about the potential repercussions of turning curation into a money-generating service. As a whole, our study provides a rich empirical contribution that focuses not just on the funding mechanisms for open data infrastructure, but also on the relationships which enable them.

\pagebreak

\section{Introduction}
Open data infrastructures (ODIs), such as research data repositories, DOI registration agencies, or federal governmental portals, underpin how open research is practiced. Despite their importance, the long-term availability of ODIs is far from guaranteed \parencite{strecker_disappearing_2023}, as persistent challenges related to technological change, geopolitical forces, and short-term financing threaten their sustainability \parencite{gregory_sustaining_2025}. While the financial mechanisms supporting open data infrastructures are particularly critical \parencite{borgman_data_2021}, they are also often complex and precarious, involving a range of funding streams and dynamic relationships with other actors in the scholarly system \parencite{skinner_2025_2025}. 

Drawing on a mixed methods study of Dryad, a well-known open data repository, we argue that understanding these infrastructural relationships, and the role of strategic business thinking within them, is critical to financial sustainability. We find that while taking a business-oriented approach can be necessary, it is also one that affects other critical factors important in an infrastructure’s longevity, namely interpretations of value(s), community, and governance. This study explores the questions: Which types of infrastructural relationships help or hinder financial sustainability? How do processes of assetization shape other critical aspects of sustaining open data infrastructures? What can other ODIs learn from the tensions which emerge when ‘thinking like a business?’

While financial security is a core concern, sustaining ODIs, or making them available for the long-term, involves many complex, inter-related factors. Tensions related to governance \parencite{mounier_sustaining_2023}; maintenance \parencite{borgman_durability_2016}, and communities and usage \parencite{fenlon_mutual_2023} all impact longevity. Sustaining infrastructure includes balancing the "infrastructural work" of managing content and keeping the lights on and the "institutional work" of building relationships with other actors \parencite{mayernik_toward_2023}. It is a constant process of balancing tensions between dynamic factors, as well as reconfiguring different forms of relationships \parencite{edwards_understanding_2007}.

Business activities, particularly “assetization” \parencite{birch_rethinking_2017}, involve negotiating relationships between actors, claims, and practices to create assets capable of generating long-term revenues. Such processes are flexibly organized within business models, which exert their own agency within an organization, shaping how relationships with consumers and investors are coordinated \parencite{doganova_what_2009}. Assetizaiton involves a specific way of thinking, where objects or services are viewed as potential financial investments. This perspective, which we term ‘thinking like a business,’ in turn influences how assets are understood and governed \parencite{birch_assetization_2024}. 

We bring these perspectives – the importance of reconfiguring relationships and business thinking – to our analysis. After situating our work in relevant literature, we describe Dryad, an open data infrastructure supporting research data sharing and reuse, and how it has reconfigured infrastructural relationships and revenue streams over time. We identify four types of relationships with customers, collaborators, and competitors that have been critical to Dryad’s financial evolution: reinforcing, forging, positioning, and excluding relationships. We then analyze how Dryad's strategic efforts to think like a business have shaped interpretations of value(s), community, and governance. We conclude by highlighting emerging tensions that provide insight for other open infrastructures working to become financially sustainable. Our analysis adds to the existing literature by focusing not just on financial mechanisms (although those emerge) but on the relationships which enable them. 

\section{Background} \label{section_2}
We situate our study in work on knowledge infrastructures, highlighting the importance of relationships and prevalence of financial risk, before examining critical literature on information economies, including assetization, and existing funding for open data infrastructures.

\subsection{Sustaining Open Infrastructures through Relationships} \label{section_2-1}
Open data infrastructures, like other knowledge infrastructures, are neither purely technical nor inherently stable \parencite{star_steps_1994,star_steps_1996}. Rather, they are best thought of as webs of sociotechnical relationships between people, technology, content, and policies \parencite{edwards_vast_2010}. ODIs are situated within broader networks and systems, where changes in one aspect can shape and affect the stability of others \parencite{edwards_understanding_2007}. This creates a constant state of evolution and reconfiguration of the social, technical, and political elements composing an infrastructure, as well as its relationships with others in the system \parencite{karasti_artful_2004,karasti_studying_2018}.

Open infrastructures (OIs) more broadly are organized around principles and values of openness such as inclusion and equity, transparent governance, and community accountability \parencite{mayer_9_2026}; they strive to make content openly available (when possible), often using open technologies \parencite{thaney_what_2025}. As \citeauthor{mayer_9_2026} summarizes, “openness does not refer only to access as an end state, but to the practical and institutional conditions under which knowledge can circulate, be reused, and be collectively stewarded.” \parencite{mayer_9_2026} Openness is mediated by “judicious connections,” carefully considered relationships between human actors that are rooted in trust and shared values \parencite{leonelli_philosophy_2023}. Within these relationships, decisions are made about what content to open, when, and to whom; these considered relationships can also be critical in sustaining open infrastructures of different types \parencite{trauttmansdorff_resilience_2026}. 

Work in infrastructure studies points to various tensions important in sustaining (open) infrastructures related to values, power, and voice \parencite{karasti_knowledge_2016}. For open infrastructures, this is visible in frictions around values of inclusivity and transparency or differences with proprietary systems \parencite{okune_decolonizing_2019,gregory_sustaining_2025}. An infrastructure’s longevity is shaped by the various communities which develop around it as it is produced, used and maintained \parencite{fenlon_mutual_2023,thomer_patchwork_2024}, as well as how it designs and implements forms of governance \parencite{mounier_sustaining_2023}. Throughout, relationships with and between different actors influence longevity, from users to curation workers to technologies \parencite{ribes_long_2009}. Cultivating funding relationships are key; the “institutional work” of engaging with policymakers and funders is necessary to help ODIs transition from project status to becoming more (financially) stable organizations \parencite{mayernik_toward_2023}.

The digital preservation literature highlights various risks which impact an ODI’s longevity. Synthesizing key studies, \citeauthor{recker_hope_2026} classify these risk factors as legal, technical, procedural, or organizational (related to economic failure) \parencite{recker_hope_2026}. Organizational factors in particular have been shown to result in the closure of ODIs. An analysis using the re3data registry identified nearly 200 data repositories that had shut down \parencite{strecker_disappearing_2023}; although the reason for most of these closures was unclear, organizational failures, i.e. the ending of project-related funding, were often a contributing factor. 

Despite these risks, financial models for sustainable ODIs remain under-researched \parencite{eschenfelder_financial_2022}, although practical sustainability-related frameworks recognize their importance. Two grass-roots frameworks have received particular attention. The Principles of Open Scholarly Infrastructure (POSI) \parencite{posi_adopters_principles_2025} provide guidelines for OIs to design governance, financial, and insurance strategies in line with the values of open infrastructure. POSI emphasizes that OIs should be transparent about finances and use revenue sources in-line with an organization’s mission (See Section \ref{section_2-2-3}). The Global Open Research Commons (GORC) \parencite{jones_global_2023,treloar_global_2025} stems from work within the Research Data Alliance to create a framework for fostering an interoperable commons of research data infrastructures. GORC emphasizes the need for long-term financial models, governance structures that can change as ODIs evolve, and the need to maintain interoperable connections with other infrastructures.

\subsection{Information Economies for Open Science: A Critical Perspective} \label{section_2-2}
There is a long history of interest in the “information economies” that can best support open science and data sharing (e.g. \cite{dalle_advancing_2005,david_can_2004}). This work is rooted in information studies and cognate fields of science and technology studies (STS) and political economy and draws on examinations of scientific knowledge more broadly (e.g. \cite{callon_is_1994,radder_which_2017}). As a whole, this literature takes a critical approach to 1) seeing openness as a public good, including related tensions between collective and transactional funding models, and 2) understanding goods/services, value, and business models as being socially constructed.

\subsubsection{From Public Goods to Tensions between Funding Models}\label{section_2-2-1}
Open information and open data are argued to be “public goods,” a special type of intangible economic good that is not depleted when other people use it (a property referred to as non-rivalry) and one that cannot be appropriated in ways that exclude its use by others (non-excludability) \parencite{david_economic_2003}. Competitive markets do not do an efficient job of managing the costs of producing and sharing public goods, because these markets usually assign costs to the consumers or users \parencite{david_economic_2003}. While commonly adopted, this perspective has been critiqued. The concept of non-rivalry, e.g., suggests that once a good has been produced, there is no longer a need for investment \parencite{callon_is_1994}. Arguably, in the case of openly available information and data, there are continued costs in making these goods available, related to storage, curation and maintenance. More recent scholarship has questioned if open publicly-funded research outputs are in fact global public goods, as they are sometimes tangible objects (e.g. a publication) rather than intangible (e.g. knowledge or ideas) and do not circulate equally across the world, due to geopolitical frictions \parencite{ma_limits_2025}. Rather than being purely public or private, the differentiation is likely a bit blurrier, with scientific knowledge exhibiting characteristics of both types of goods \parencite{radder_which_2017}. 

Public goods are related to the idea of the commons \parencite{radder_which_2017}, which is prevalent within open scholarly communication \parencite{moore_care-full_2018}. Rooted in ideas about shared natural resources, “the commons” is not just about having freely accessible resources, but is rather about how those resources are produced, managed, and governed \parencite{moore_care-full_2018,hess_understanding_2006}. At a high level, the commons view argues that collective ownership is preferable to private ownership for something with public good characteristics \parencite{radder_which_2017}. 

The idea of collective ownership is reflected in collective funding models in scholarly communication, such as membership or consortia models for open access publishing. Such collective models are positioned in opposition to transactional models, where money is exchanged for a single use, as in paying an article processing charge (APC) to make an article open access \parencite{mellins-cohen_classifying_2024}, or a data publication charge for an individual dataset. \citeauthor{chodacki_not_2026} argues that collective rather than transactional models best match the shared benefit that open data infrastructures provide across the scholarly system \parencite{chodacki_not_2026}.

Another, related tension is between views about proprietary organizations, such as large, for-profit publishers, and more commons-centric, often non-profit, companies \parencite{ma_limits_2025,moore_publishing_2025}. These different organizations are regularly framed as having competing logics, with the former focusing on market logics designed to achieve financial return and the latter on community- and service-oriented values \parencite{moore_publishing_2025}. Such distinctions tend not to be that binary in practice, as organizations are rarely purely profit- or mission-oriented, but are hybrid in nature, where different logics can exist at the same time \parencite{malhotra_hybridity_2025}.

\subsubsection{Co-construction: Assetization and Business Models}\label{section_2-2-2}
STS-related scholarship emphasizes the co-constructed nature of goods and services; determining value; and business models. Michel \citeauthor{callon_economy_2002} argue that goods are made through ongoing processes of qualifying and re-qualifying products, where consumers and producers together define “what the [goods] are, what they do, and what value they hold.” \parencite{callon_economy_2002} Qualification is particularly visible in service-providing activities in information economies, where the value of a service is created through interactions between providers and customers. Firms within such economies compete by differentiating the value of goods and services through these processes, and by building and breaking relationships with customers, as people become attached or detached from particular offerings \parencite{callon_economy_2002}.

Qualification informs what Kean Birch terms “assetization,” processes through which organizations turn goods or services into assets that generate durable rents rather than one-time revenues \parencite{birch_rethinking_2017}. The concept of assetization is particularly visible in our material and is one which we draw on in our analysis. Assetization occurs when firms need to rethink what gets valued and how, which in turn requires “an examination of the coproduction of these practices with organizational forms and their governance” \parencite{birch_rethinking_2017}. Assetization shapes and transforms a particular good or service through its very framing as an asset; it involves a particular way of thinking, of seeing things as (potential) financial investments and then relating to or governing them through this perspective \parencite{birch_assetization_2024,muniesa_time_2020}. 

Business models, flexible combinations of activities centered on creating and capturing value \parencite{zott_business_2010}, are also not neutral but have their own agency. How they are performed in practice shapes how organizations structure and coordinate work and relationships with other actors, such as investors and consumers, but also internally \parencite{doganova_what_2009,muniesa_time_2020}. Business models therefore work as “calculative and narrative devices” \parencite{doganova_what_2009}, performing markets and framing constitutive activities such as assetization \parencite{birch_rethinking_2017}. 

\subsubsection{Funding Open Data Infrastructure: Revenue Streams and Costs in Practice} \label{section_2-2-3}
ODIs need to put this into practice. A foundational report from the \citeauthor{organisation_for_economic_co-operation_and_developmentoecd_business_2017} argues that ODIs should explicitly create business models and highlights the types of activities that should be included: identifying and articulating value propositions, potential stakeholders, revenue sources, and costs of data sharing and curation \parencite{organisation_for_economic_co-operation_and_developmentoecd_business_2017}. Underscoring the importance of flexibility, business models should support both large scale change as well as minute adjustments in revenue streams \parencite{eschenfelder_financial_2022}.

ODIs typically acquire revenue through different mechanisms, i.e. institutional support, grants, or fees \parencite{eschenfelder_financial_2022,ge_gorman_and_professor_jennifer_rowley_funding_2015}. Public funding is often the most common revenue source for ODIs, \parencite{attwood_longevity_2015,skinner_2025_2025}. This funding is usually distributed through (multiple), short-term grants from government organizations, reflecting broader structures for funding scientific research \parencite{imker_who_2020}. These short funding cycles are not ideal for ODIs, as they encourage relying on short-term funds to cover longer-term costs, which can make organizations fragile \parencite{posi_adopters_principles_2025}. 

Revenue sources for ODIs should also be in line with open principles and values \parencite{posi_adopters_principles_2025,shankar_sustaining_2015}. The OECD report advocates that charging data depositors and charging for services (which do not exhibit characteristics of public goods) best supports open values and equity of access \parencite{organisation_for_economic_co-operation_and_developmentoecd_business_2017}. Despite this advice, selling services has been relatively rare for data repositories \parencite{eschenfelder_financial_2022}, and other pragmatic factors, such as timelines, funding eligibility, and legal restrictions shape an ODI’s mix of revenue streams \parencite{hooft_financing_2025}.

The literature focuses more on potential revenue sources and less on the actual costs of data sharing and curation, which can be related to characteristics of collections, or repository functions \parencite{national_academies_of_sciences_engineering_and_medicine_life-cycle_2020}. Costs typically include a mixture of fixed costs (e.g. salaries, rents) and variable costs \parencite{lhours_cost_2014}. Costs for staff typically exceed costs for hardware \parencite{mohr_making_2024} and are determined by complex factors, i.e. the types of tasks to be performed, their duration, and required skills \parencite{national_academies_of_sciences_engineering_and_medicine_life-cycle_2020}. Although there are several models for determining the cost of digital curation \parencite{bogvad_kejser_evaluation_2014}, these “soft costs” can be difficult to precisely determine \parencite{borgman_data_2025}. They are, however, significant, with some indication that costs for data curation, on average, exceed costs for system maintenance and development, preservation, and data access \parencite{organisation_for_economic_co-operation_and_developmentoecd_business_2017}.

\subsubsection{Summary}\label{section_2-2-4}
Overall, the literature indicates that financially sustaining ODIs is complex and involves relationship work of different sorts: managing dynamic relations with other actors and infrastructures, which shape organizational risk, as well as managing relationships within processes of qualification and assetization. Business models include and shape these processes as they are put into practice. For ODIs, these include activities such as negotiating and determining the value of goods and services, searching for revenue streams matching open values; determining costs for data sharing; and turning data or services into assets capable of generating recurring value \parencite{birch_rethinking_2017}. We bring these perspectives – particularly the importance of relationships in sustaining infrastructure and business thinking – to explore how Dryad has worked towards financial sustainability over time.

\section{Methods} \label{section_3}
To explore Dryad’s (financial) evolution, we designed a mixed methods study combining semi-structured interviews, document and material review, and a quantitative descriptive analysis of Dryad’s data and data references.

\subsection{Interviews and document review} \label{section_3-1}
We conducted 14 one-hour, online semi-structured interviews between April and August 2025 with current and past Dryad staff and members of the Board of Directors. Interviews lasted approximately 60 minutes and were transcribed for analysis. To recruit participants, a relationship was first developed with Dryad senior staff, who agreed to support the study. Participants were then recruited based on recommendations and snowball sampling. Recruitment stopped once we had spoken with representatives from all professional roles at Dryad who were willing to participate in this study. In sum, we spoke with two full-time curators and technologists, eight board members (split equally between publishing and institutional representatives), and three (former) staff. We also spoke with one employee at another OI with extensive experience of Dryad. Speaking with these individuals enabled us to collect enough data to answer our research questions and achieve sufficiency in our data \parencite{ladonna_beyond_2021,vasileiou_characterising_2018}. We describe demographics at an aggregate level to protect the anonymity of participants and refer to participants by number (PX) in the findings. 

Participants were invited to bring a document to serve as a discussion prompt; these included blogposts, reports, and policy documents. Participants explained the importance of these prompts and their relationship to sustainability. Other interview questions addressed different factors influencing the sustainability of ODIs, identified from the literature, e.g. funding, governance, communities, and relationships to other repositories. Questions were tailored to match participants’ expertise. ATLAS.ti \parencite{noauthor_atlasti_2022} was used to support a combination of deductive and inductive coding and reflexive thematic analysis \parencite{braun_using_2006,braun_reflecting_2019}.

We developed the codebook iteratively in rounds of discussion with both authors. After initial deductive code development, both authors coded a sample of three transcripts; codes were then further discussed. The authors then re-coded the set of transcripts and had a further discussion. The first author then coded the remaining transcripts; the second author read the remaining transcripts to feed into further discussion and iteration. Deductive codes were based on literature about sustaining knowledge and data infrastructures (Section \ref{section_2-1}) as well as the interview protocol. Deductive codes related to governance (e.g. community governance; board organization); funding (e.g. descriptions of the new fee structure); Dryad’s relation to other repositories; and community needs and engagement. Inductive codes included reconfigurations of relationships and business thinking. These inductive codes informed the framing of our analysis. The interview protocol and codebook are openly available, along with our quantitative data \parencite{strecker_data_2026}. This study is part of a project that has received ethical approval from Leiden University.

We also closely read openly available documentation about Dryad. This material included blogposts and news items written by Dryad staff, organizational bylaws from 2016 and 2024, a selection of minutes from board meetings and annual reports, and information on Dryad’s website and related websites. We also attended two open webinars in spring 2025 for publishers and institutions explaining the new fee structure. This material was not systematically coded, but informed conversations with interviewees and nuanced our understanding of Dryad’s development over time. 

\subsection{Descriptive data analysis} \label{section_3-2}
We conducted a descriptive, quantitative analysis of Dryad’s metadata to characterize the repository and understand the use of Dryad’s data. This analysis supports and adds further detail to the findings emerging from our qualitative methods. We further analyzed Dryad’s federal funding over time using an openly available dataset (Riordan et al., 2025) to understand the contribution of project funding to Dryad’s financial sustainability. 

As a first step, we retrieved DOIs from the Dryad API on 2025-06-24 . At that time, 64.774 DOIs were accessible through the API. Metadata of datasets (author last names, publication year, publisher name and resource type) were added via the DataCite API on 2025-07-11 . The resulting sample included metadata for 64.774 Dryad datasets. We were also interested in understanding more about the use of Dryad’s data, as indicated by citations. There are several sources that capture citations of datasets, relying on different strategies to identify data citations. To ensure that publications referencing Dryad data were covered as completely as possible, three sources were combined (Table \ref{tab:sources}). These sources were searched for the 64.774 DOIs retrieved from the Dryad API in the previous step.
\begin{table}[H]
    \centering
    \begin{tabular}{|p{4cm}|p{2cm}|p{4cm}|p{2.5cm}|} \hline
         source & date & method & number of citations \\ \hline \hline
         Data Citation Corpus v3.0 \parencite{datacite_data_2025} & 2025-02-01 & Dryad DOI strings were matched with the "dataset" column & 46.873 \\ \hline
         DataCite REST API - relationships section\footnote{DataCite - Consuming Citations and References: \url{https://support.datacite.org/docs/consuming-citations-and-references}; last accessed: 2026-03-04} & 2025-07-11 & “citations” were retrieved from the relationships section & 54.335 \\ \hline
         OpenAlex\footnote{OpenAlex API documentation: \url{https://docs.openalex.org/how-to-use-the-api/api-overview}; last accessed: 2026-03-04} & 2025-07-04 & Dryad DOIs were passed as an argument to the "cites" filter & 492 \\ \hline
    \end{tabular}
    \caption{Overview of sources used to retrieve references to Dryad datasets.}
    \label{tab:sources}
\end{table}

References were deduplicated. DOIs of all publications referencing Dryad data were checked; 318 were removed because the DOI did not resolve. This resulted in 50.126 references in total. Similar to previous studies, the type of relationship between the Dryad dataset and the referencing publication was determined by comparing author last names \parencite{cohen_matching_2026}. Following \citeauthor{bobrov_typology_2026}, if there was overlap in author lists, the relationship was categorized as “data use” and “data reuse” if there was not \parencite{bobrov_typology_2026}. The comparison was based on author last names, due to limited availability of author identifiers. As last names may not uniquely identify authors, it is possible that this approach led to underestimating data reuse. Because the publisher field allows free text, the values are very heterogenous. Therefore, publisher names of referencing publications were harmonized and checked. Mentions of metadata aggregators and missing values were corrected. Publisher imprints were then grouped according to \citeauthor{haupka_openbib_2025} \parencite{haupka_openbib_2025}. This allowed us to group referencing publications at the company level, matching descriptions of publishing partners in Dryad’s documentation.

For subject information, only labels assigned from the OECD Fields of Science Classification (FOS) were considered, as they were the most widely used controlled vocabulary in the sample. These labels were aggregated at the highest level of the classification. Each label was counted once per dataset. In total, 24.305 FOS labels of 24.148 datasets were included. To analyze Dryad’s federal funding, data from the 2025 State of Open Infrastructure Report \parencite{riordan_data_2025} was reused. Eight direct grants awarded to Dryad with information about start and end date were included. The total amount for each of these awards was divided by the number of project months. 
These approaches allowed us to characterize Dryad and its data, analyze its federal funding over time, and support the moments of relationship reconfiguration revealed through our qualitative methods.

\section{Findings} \label{section_4}
We begin by describing Dryad and the evolution of its financial model by tracing changes in key relationships with other actors. We then explore how the strategic development of Dryad’s new business model and process of assetization evoked changes in various elements important in sustainability: namely, interpretations of value(s), notions of community, and governance.

\subsection{Introducing Dryad} \label{section_4-1}
\begin{figure}[H]
    \centering
    \includegraphics[width=1\linewidth]{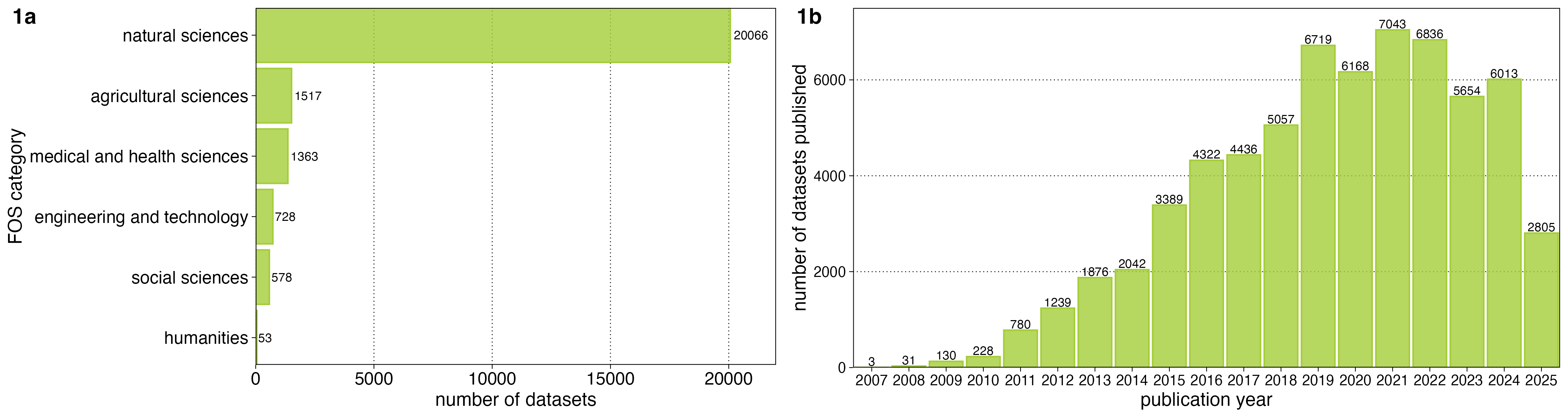}
    \caption{Characterization of datasets in Dryad. Figure 1a. Subject classification of datasets in sample according to FOS categories. Figure 1b. Number of datasets in Dryad over time.}
    \label{fig:figure_1}
\end{figure}

Dryad is an “open data platform and a community committed to the open availability and routine re-use of all research data” \parencite{dryad_who_nodate}. Originally rooted in the life sciences, Dryad portrays itself as a generalist data repository that accepts data regardless of type, format, or disciplinary focus \parencite{us_national_library_of_medicine_about_nodate}. Despite this, the majority of datasets in our analysis continue to be related to the life sciences (Figure \ref{fig:figure_1} a), with 74.5 \% of datasets in the largest subject category, natural sciences, coming from the biological sciences. This category continues to see the highest number of deposits each year. One participant emphasized that being a generalist repository is not just related to content, but also to the type of (more general) metadata and curation activities provided. Participants mentioned that becoming a generalist repository was a business decision that allowed Dryad to broaden its potential customer base.

\subsection{Reconfiguring Relationships for Sustainability}\label{section_4-2}
\begin{figure}[H]
    \centering
    \includegraphics[width=1\linewidth]{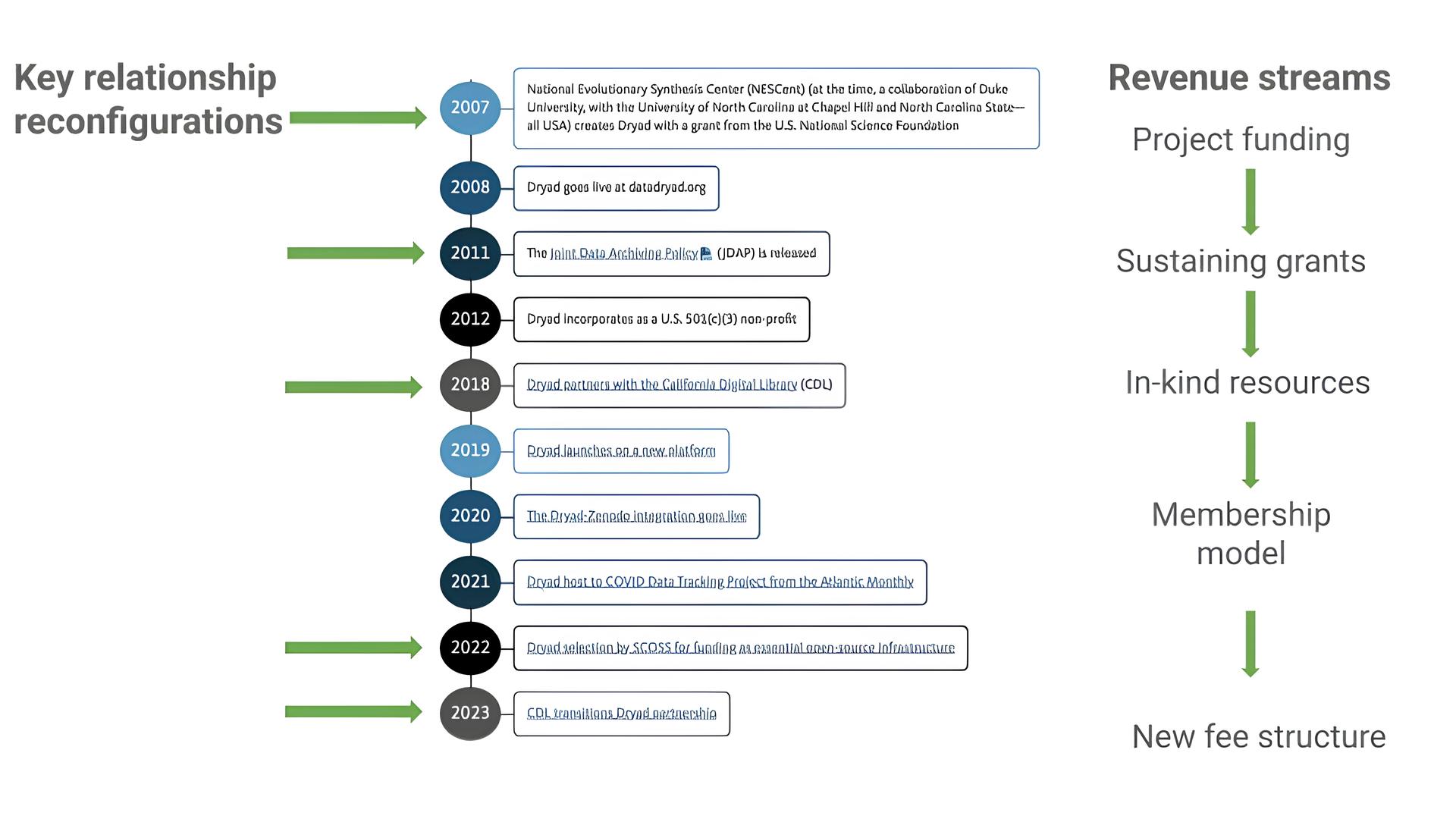}
    \caption{Key moments of relationship and revenue reconfigurations over time. The timeline was created from a screenshot on Dryad’s webpage (December, 2025). Annotations to the timeline were added by the authors.}
    \label{fig:figure_2}
\end{figure}

\begin{figure}[H]
    \centering
    \includegraphics[width=1\linewidth]{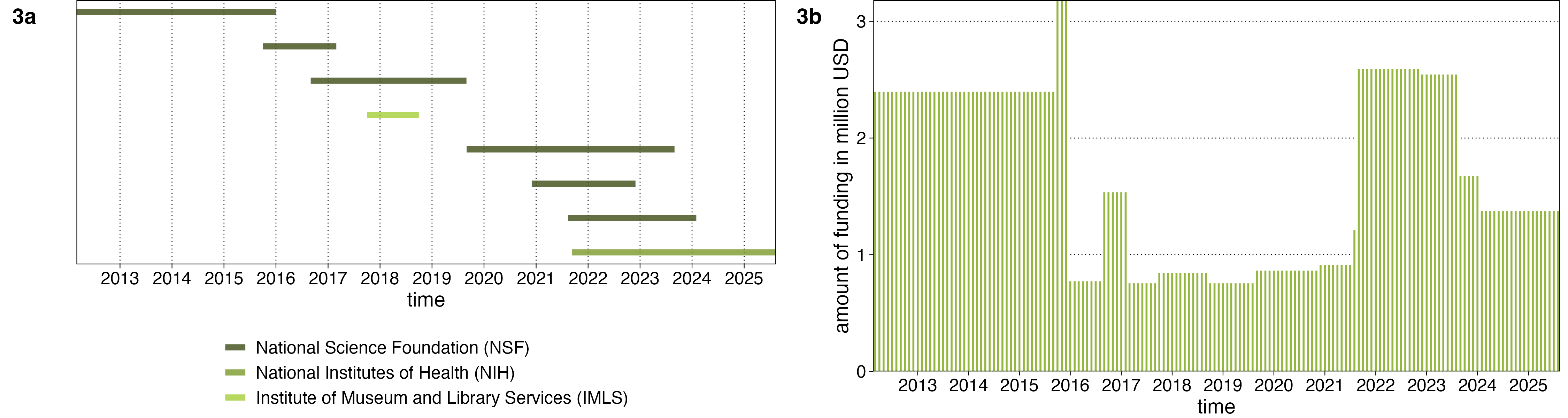}
    \caption{Federal funding for Dryad over time. Figure 3a. Funding streams according to funding agency. Figure 3b. Combined funding amounts across funders}
    \label{fig:figure_3}
\end{figure}

Figure \ref{fig:figure_1} b demonstrates an overall growth in the number of deposited datasets over time. There have been moments where growth plateaued, reflecting a slower pace of research during COVID-19, as well as key moments in Dryad’s history. Many of these key moments were related to reconfigurations in relationships and revenue streams (Figure \ref{fig:figure_2}). Figure \ref{fig:figure_2} demonstrates a common pattern in the development of ODIs: transitioning from a time-limited project to institutional embedding to becoming an independent organization \parencite{organisation_for_economic_co-operation_and_developmentoecd_business_2017}.
The first important relationship configuration in Figure 2 describes the repository’s origin. Dryad has its roots in a group of researchers at the National Evolutionary Synthesis Center, which was funded by the United States National Science Foundation (NSF). These researchers proactively created the first version of Dryad to meet publishers’ and funders’ data sharing policies. Initial project funding came from the researchers’ institutions as well as dedicated grants from the NSF. The grassroots nature of these beginnings resonated with the broader OI community:
\begin{quote}
    [There were] other people who are in the story, working at other places who saw Dryad as this exemplar of the community rising up to meet the challenges of the day. You know, data policies [..] were all taking hold and people were like, “How do you address it?” And Dryad, as an example of researchers, saying, “Let’s just build something ourselves for ourselves,” was always a cool idea. (P7)
\end{quote}
The development of Dryad reflects the complexity of funding streams for ODIs. This complexity is also visible in federal funding over time. Project funding was an important revenue source throughout Dryad’s history, with a particular reliance on NSF grants (Figure \ref{fig:figure_3} a). Although consistently present, the amount of monies from federal grants has varied (Figure \ref{fig:figure_3} b), with a notable drop corresponding roughly to Dryad’s relationship with the California Digital Library (Section \ref{section_4-2-2}).
Our interview data demonstrate different ways in which Dyrad has reconfigured relationships with other actors, both customers as well as (potential) collaborators and competitors, as it has worked to sustain itself. We classify these as \emph{reinforcing}; \emph{forging}; \emph{positioning}; and \emph{excluding} relationships. By nature dynamic and overlapping, this classification provides insight into the relationships supporting ODIs over time.

\subsubsection{Reinforcing relationships with customers (publishers)} \label{section_4-2-1}
\begin{figure}[H]
    \centering
    \includegraphics[width=1\linewidth]{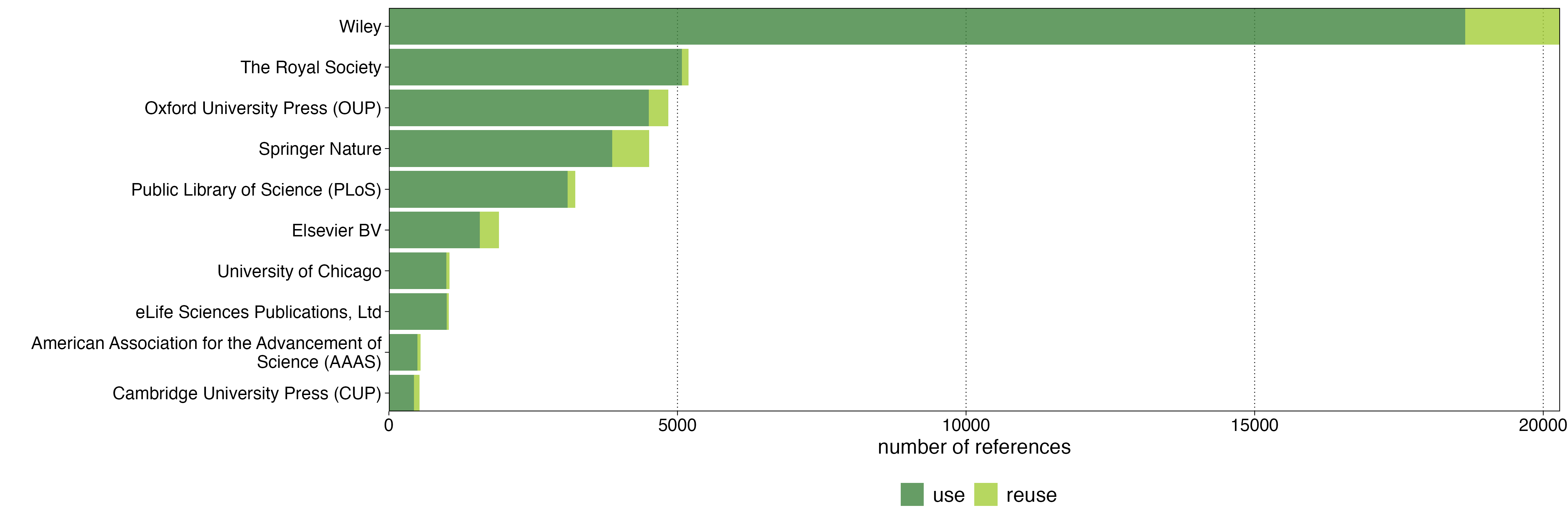}
    \caption{Number of references classified as use and reuse for the top ten publishers in our sample. Use indicates that the datasets and publications were classified as being created by the same author. Reuse indicates that a dataset and publication were created by different individuals.}
    \label{fig:figure_4}
\end{figure}

Strong relationships with publishers have been key to Dryad's persistence. Seeing the need to make data supporting publications available, but also lacking a viable way to do so, publishers encouraged the development and use of Dryad. In 2011, e.g., the Joint Data Archiving Policy (JDAP) was adopted in a coordinated fashion by many leading journals to require data sharing for published articles. Many journals recommended the use of Dryad in their statements. Notably, the policy was not released until Dryad was operational. As one participant described: 
\begin{quote}
    There were two things that happened; the center obtained additional funding from the National Science Foundation specifically to build Dryad. And then there was a coalition of these journals that said: as soon as Dryad is ready to go, the journals will start requiring that all of our articles have associated data published. And the journals simultaneously launched this policy [..] because they didn’t want any one journal to suddenly launch this new strict policy and then have all the authors flee to the other journals. (P5)
\end{quote}
This relationship with publishers became more seamless with time, as publishers embedded Dryad in their workflows. Technical integrations with publishing systems, such as automatic metadata completion for deposited data, have strengthened this relationship. This embedding is also visible in the citation of Dryad datasets in published articles; 69.18 \% of Dryad datasets have been cited at least once. The vast majority (93.6 \%) of data citations classify as instances of “use”, meaning that the datasets were cited by data creators. This pattern holds across the top publishers in our sample (Figure \ref{fig:figure_4}), which includes five publishers listed as “founding partners” on the Dryad website: Wiley, the Royal Society, Oxford University Press, PLOS, and AAS, demonstrating the longevity of these publishing relationships. 

\subsubsection{Forging relationships with customers (institutions)} \label{section_4-2-2}
In addition to reinforcing relationships, Dryad forged new relationships, most notably with institutions. Some of these relationships were carefully cultivated to reach new institutional customers; some were more serendipitous, as in the case of a core relationship in Dryad’s history: its partnership with the California Digital Library (CDL). This partnership began at a moment when Dryad was struggling to make ends meet, in part because of technical limitations. At the same time, CDL was developing an institutional repository for data that felt like a siloed and under-used resource. Informal, honest conversations about the struggles of funding open infrastructure created an environment for “radical collaboration:”
\begin{quote}
    Open infrastructures succeed when you—without any ego or pretense, just can open your doors to the opportunities that come with openness. So, really thinking not just about planning for openness, but actually just being there for serendipitous or opportunistic openness and collaboration. (P8)
\end{quote}
This new collaboration was key for Dryad’s survival, as suggested by the increase in deposited datasets at this time (Figure \ref{fig:figure_1} b). CDL provided in-kind contributions in the form of dedicated staff for technical development, product management, outreach, and strategy. These in-kind contributions were seen as critical for moving Dryad to a new technical platform and expanding the existing business model.
\begin{quote}
    We weren’t just talking about technology. We were also talking about business models. From the very beginning, we were saying, “For this to work, Dryad needs to change. Not just its product and its platform, but also its offering.” [..] Before they did not publish datasets, unless they were attached to a journal article. And we were like, from an institutional perspective, we have datasets that don’t have that [..] We would need features—administrative features—that would help with institutional use cases. And that kind of a thing meant that they had to agree to change from a business model before CDL even signed. (P8)
\end{quote}
The relationship with CDL, based on a shared sense of values, widened Dryad’s reach to enroll other institutional members.
\begin{quote}
    Institutions were key to Dryad’s success because institutions shared values with what Dryad was presenting. When CDL convinced everybody to take this path and make this leap and say, “Why don’t we just work together and [..] all ships will rise,” the thing that made it happen was saying, “We share values. Here are our institutional values and here are Dryad’s values. They are the same.” I think that that’s also a piece of this puzzle [..] libraries and institutions and open infrastructure, they share the same values. (P7)
\end{quote}

\subsubsection{Positioning relationships with collaborators and competitors} \label{section_4-2-3}
\begin{figure}[H]
    \centering
    \includegraphics[width=1\linewidth]{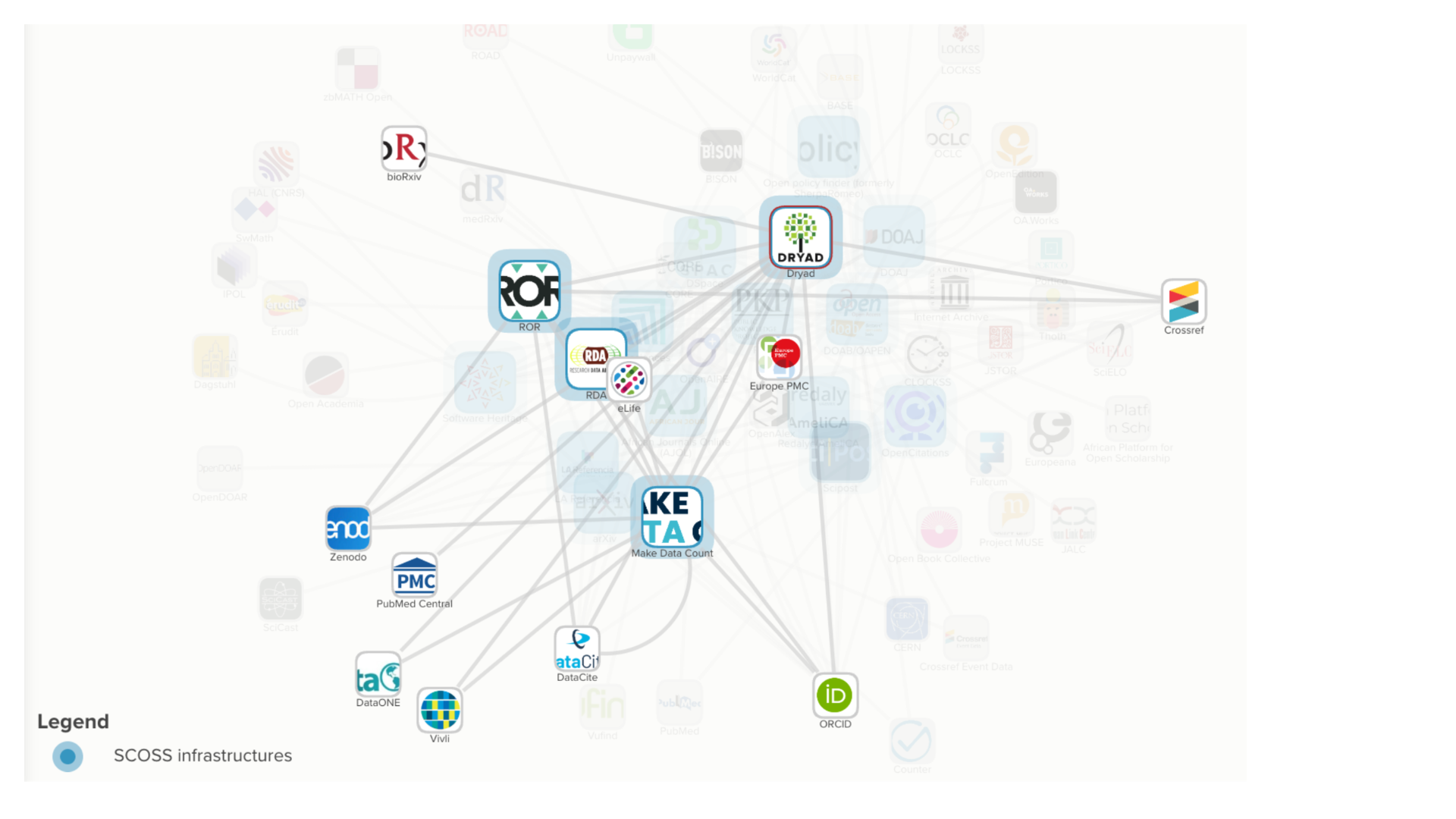}
    \caption{Technical and project integrations between Dryad and other SCOSS “family members”. Map retrieved February 11, 2026 from \url{https://scoss.org/what-is-scoss/scossfamily/}}
    \label{fig:figure_5}
\end{figure}

Dryad has also engaged in positioning relationships, where it positions itself alongside or differentiates itself from certain infrastructures, particularly data repositories, scholarly communication infrastructures, and open/proprietary infrastructures.
\vspace{0.3cm}
\\
\textbf{Positioning with data repositories }
\\
Dryad's positioning varies in relation to different types of data repositories, be those disciplinary, institutional, or other generalist repositories. Numerous participants reiterated the importance of having repositories with specific metadata and curation workflows for data from particular disciplines. Dryad positions itself as complementing these disciplinary repositories, as a place for heterogeneous data. In contrast, Dryad clearly positions itself as a replacement for institutional repositories. Providing curatorial and storage services for institutions is seen as a resource-efficient way to help institutions meet the needs of their researchers (Section \ref{section_4-2-1}). At the same time, other generalist repositories, such as Zenodo or Dataverse, are seen as “peers.”
\begin{quote}
    I think that many people [..] would say, “Do we really need Dryad and Zenodo if they’re both open or something like that?” When in reality, it’s not about that. What the question in that case is, why do we have all these data institutional repositories when we have things like Dryad and Zenodo? [..] Why are there thousands of [institutional] data repositories around the world with ten datasets in them? (P7)
\end{quote}
\textbf{Positioning with Other Scholarly Communication Infrastructures}
\\
Dryad has engaged in the institutional, project, and policy work necessary to embed itself within the broader scholarly ecosystem. In 2022, Dryad became part of the Global Sustainability Coalition for Open Science Services’ (SCOSS) list of essential infrastructures, which came with extra funding and facilitated connections with other SCOSS “family members” (Figure \ref{fig:figure_5}). This map depicts technical integrations and joint projects between Dryad and generalist data repositories, societies, persistent identifier providers, and publishing organizations. These embeddings anchor Dryad in a web of related actors across the scholarly system. 
\textbf{Positioning with Open/Proprietary Infrastructures}
\\
Dryad carefully positions itself as an open infrastructure, often in opposition to proprietary or commercial infrastructures. Openness is framed in different ways, ranging from publishing all data openly using a CC0-licence to being transparent about governance documents, to having a “nonprofit ethos.” 
\begin{quote}
    I think of all of the generalist repositories, Dryad is the one that's most attractive personally, because they do have the non-profit ethos of “we're doing this to support good science and advance research,” not to make a buck. And certainly, they need to make a buck, and they want to make a buck. But that's to sustain the open infrastructure and the services not to go retire on the Riviera or whatever else everyone does. (P2)
\end{quote}
Openness of technology is also seen as a differentiating factor between Dryad and both proprietary infrastructures as well as other nonprofit organizations. 
\begin{quote}
    So, it’s not only that we’re nonprofit [..], it’s the operation of the technology. My understanding is that open infrastructures are not necessarily run by open organizations. Hopefully they are, because the values are consistent in that way, but open infrastructure is where the software is open itself [..] And that stands in contrast to a commercial organization or another nonprofit organization that might keep its code closed. (P1)
\end{quote}
The previous quote suggests that the line between open and non-open is not always clear cut but exists instead on a spectrum. It also highlights the importance of “open values” (Section \ref{section_4-3-1}) in how Dryad positions itself; many participants see these values as a necessary differentiating factor which enables Dryad to compete against proprietary infrastructures with more resources.
\begin{quote}
    It’s sort of in competition with private publishers, where you have some private organizations that are quite big [..] And they can operate in this space because they have a very different fiscal model [..] And there’s a danger of private interests sort of out-competing public repositories [..] [but] a lot of people do value the fact that we’re not a private company that has private interests to just provide a product, but we actually really care about the values of openness (P10)
\end{quote}
In its positioning relationships, Dryad faces questions about competition and collaboration. Dryad positions itself as a competitor to institutional repositories or in opposition to proprietary infrastructure. In other relationships, collaboration and partnership are seen as a strength, e.g. in thinking about how to maximize limited financial resources within the OI community.
\begin{quote}
    This is a conversation [..] about the multiplicity of open infrastructure, all knocking on the same funders’ doors to get money. How can we all either share resources or come up with a package that says, “This is the support that open infrastructures can provide to you as an academic institution.” And how do we collectively create those synergies so that people don’t go and pick a proprietary software? (P6)
\end{quote}
Sometimes, as in the General Repository Ecosystem Initiative (GREI), competing repositories collaborate “below the value line” in a coopetition model \parencite{kim_coopetition_2020}. This promotes base level standards in the broader ecosystem without compromising (or revealing) the unique value-add of an individual repository. 
\begin{quote}
    What our experience has been is folks keep close to the vest, whatever they keep close to the vest, but we are able to engineer productive collaboration around getting all of these platforms to adopt DataCite as the one metadata schema. That hadn’t been done before [..] There’s broad agreement that that stuff is not going to win business for one repository over another (P1)
\end{quote}
Participation in GREI positioned Dryad among a select group of infrastructures. This allowed it to deepen relationships with these organizations and served as a reputational signal that it was seen to be a leading player in the open data infrastructure landscape.

\subsubsection{Excluding relationships with collaborators and competitors} \label{section_4-2-4}
Dryad makes considered choices about the types of relationships to pursue but also about those to avoid or end. Participants described avoiding relationships with organizations who had been purchased by proprietary firms, for example. Another example is the conclusion of Dryad’s relationship with the California Digital Library. Frictions concerning differing visions of financial models contributed to the ending of this relationship. Despite these frictions, both organizations (who are prominent in the OI community) saw value in concluding the relationship in an amicable way.
\begin{quote}
    We took as an exec group the decision that we need to change the model to actually get to a model that would be sustainable. And so that meant actually having costs that were effectively metered by usage [..] Now, we did have a strong disagreement on the sort of philosophical direction of that with our partners at CDL. [...] We got to an amicable place where they said, “OK, we understand what you’re doing. We would prefer you to try to continue to work in this way. But if you want to operate independently, then how do we make that work in a way that can take both organizations forward?” (P11)
\end{quote}

\subsection{Thinking Like a Business: Shaping Value(s), Community, and Governance} \label{section_4-3}
The previous section demonstrated how reconfigurations of relationships with customers, collaborators and competitors have contributed to Dryad’s financial sustainability. We now explore how thinking like a business, by implementing a new business model and engaging in a process of assetization, affected not only financial sustainability, but also other elements important in sustaining infrastructure: value(s), community and governance.

Developing Dryad’s new fee structure (Box 1) grounded this process of business thinking. The new fee model consists of three primary parts: a data publishing charge to cover the variable costs of data curation and preservation; an annual service fee to cover fixed operational costs (e.g. training, administration, and leadership salaries); and an extra fee for large datasets. These fees are paid by Dryad’s customers: institutions (i.e. universities) or publishers/societies. The model differentiates between sponsored authors, from affiliated institutions or publishers, and unsponsored authors who pay individually or are granted a fee waiver. In cases where an author is affiliated with both a publisher and an institution, the institutional members are first asked for payment.

\begin{figure}[H]
    \centering
    \includegraphics[width=1\linewidth]{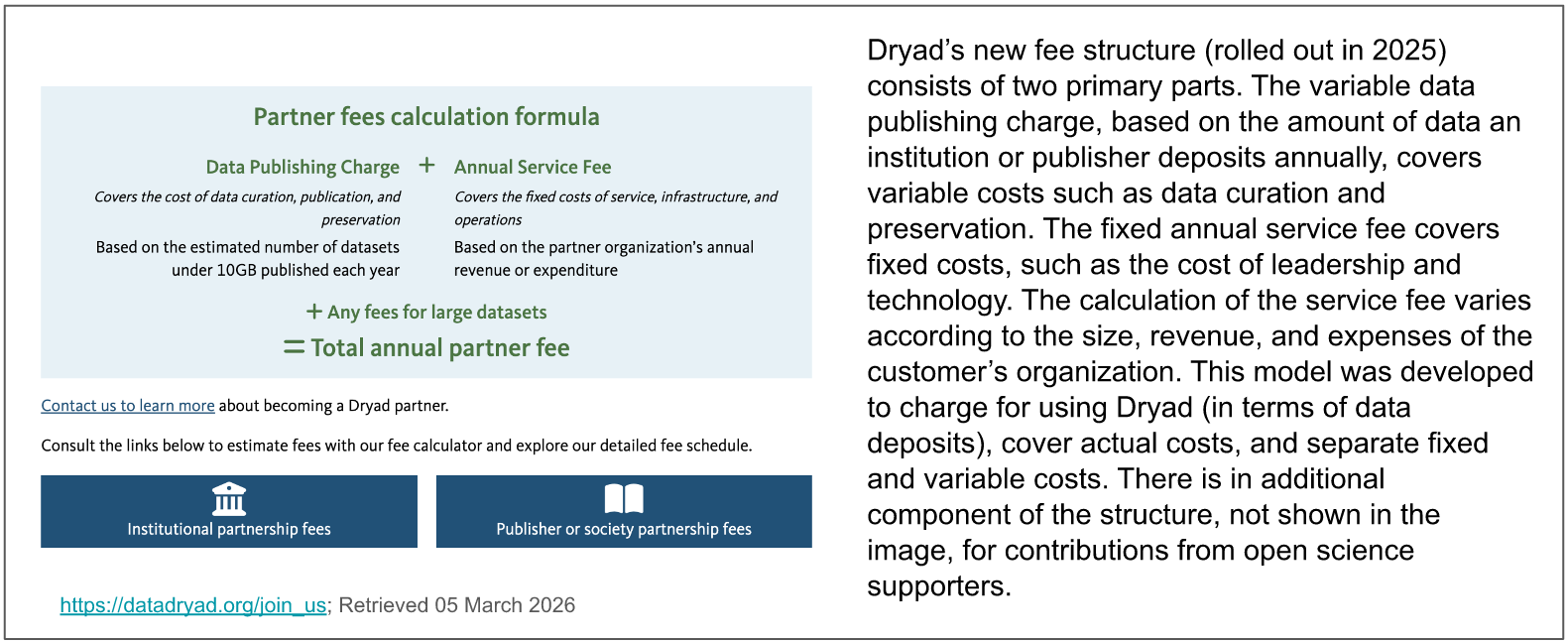}
    \caption{Description of Dryad’s new fee structure. Image is a screenshot from Dryad’s website in March 2026.}
    \label{fig:figure_6}
\end{figure}
The process of developing this structure involved a three-year period of market research and consultation centering on articulating a value proposition. While taking a business-focused approach was seen to be necessary for Dryad’s future, it was not something that came naturally to all actors: 
\begin{quote}
    I think a unique thing that sort of happens in science [..] is that most scientists, including—I will still say myself, are not business people —like, not wanting to come in and look at numbers and books. (P10)
\end{quote}
Dryad drew on the expertise of board members, including those with publishing backgrounds, a special financial task force composed of stakeholder representatives, and a financial consultant to re-evaluate Dryad’s product market fit. It aimed to develop a fee structure meeting stakeholders’ diverse interests that would provide a stable revenue source, or rent, and support principles for open infrastructure.

\subsubsection{Re-interpreting value(s)} \label{section_4-3-1}
This process involved interviews with stakeholders (customers), competitive analyses, literature reviews, and open consultations to understand and articulate the value that Dryad offered and could offer. Defining the product value of the repository was difficult; in part because of the immaturity of the “business case” for data sharing. Dryad’s new value proposition ultimately coalesced around two key points: the idea of operating “for the community” and shifting from being a platform to a “service provider.”
\begin{quote}
     Until we completed the value proposition work last year, the spiel was that Dryad is an open data publishing platform. [..] [Now] We want to be more clear that we are a nonprofit organization operating in the community interest, that we are a community-owned organization as well as being nonprofit. [..]. And the other thing we want to better articulate about our value proposition is that we are a service provider. We will operate in the community interest, but we are a trusted and reliable service provider that makes sure that data is prepared and shared and published and preserved appropriately. (P1)
\end{quote}
The framing of this service varies for different stakeholders. For publishers, the service is sold as a way to easily support data sharing mandates. For institutions, the framing focuses on providing data curation. Charging for these services involved determining and attaching monetary value to these activities, effectively turning them into assets to generate longer term income. Assigning monetary value to curation services was highly iterative, involving multiple rounds of negotiation with institutional stakeholders to develop different pricing tiers based on research intensity of a university. Even though many admitted that the curation Dryad provides is relatively lightweight, it was also seen as being a valuable service worth payment.
\begin{quote}
    Curation as an activity is very expensive [..] We at the libraries did not take that on, and so the amount of curation that Dryad provides – which I know is still minimal – is extremely attractive, because it means you don't have to have people on staff who are opening data sets and taking a look and reading and processing and those kinds of things. And even just the cursory sort of does the data match? That's really important [..] So having someone who includes in their publishing fee that service was really attractive. (P4)
\end{quote}
Participants mentioned multiple “unique selling points” (USPs), above all curation and openness. Various tensions arise around these USPs. While curation is both an asset and USP, it is carried out by a very small staff, involving part-time employees located in different countries. Some participants mentioned the potential of AI tools to take over basic curatorial tasks and reduce labor costs. As one curator explained, it is not just the curation that is of value, but the conversations and connections with researchers enabled through curation activities. A key point of value is therefore human connection, in addition to curation.

The openness of Dryad’s data and its source code were seen to be another USP. One participant discussed a tension between the perceived value of openness and convincing customers to pay for open infrastructure.  
\begin{quote}
    A challenge that we’ve had at Dryad has been how institutions – picking again on the academic institutions case – very often see the value of organizations like Dryad and the service we provide. And they really appreciate that we’re an open infrastructure [...] that we’re community-led, and that we’re not for profit [..] And in many cases, they appreciate the need for organizations like ours to be financially sustainable over time. But there is always a tension between perceived value and the money that people are willing to invest in open infrastructure. [..] I think ‘openness’ can help with the philosophical cause for open infrastructure, but at the practical level, it’s often a barrier. (P6).
\end{quote}
While openness is seen as a USP and important value of Dryad, it also complicated the development of the fee structure. This is apparent in the above quotation and in decisions about how to generate income when code and content are openly available. This tension informed the development of the value proposition – where Dryad is a provider of services rather than content – and foregrounds the importance of human support and connection.
\begin{quote}
    Certainly, it is more challenging for Dryad to have certain methods of funding because our code is open [..] So, we really center a lot of our charges in our business model on the idea of our hands-on human staff that helps people through the process. And so, everything we write in the software is in support of that, both to help the content be more well-structured before it even gets to our humans and for our humans to be able to deal with stuff after they receive it. (P5)
\end{quote}
This section demonstrates the process of assetizing curation services and how the fee structure reflects certain values. Curation is associated with a value of care and human connection; and openness with transparency and accessibility. Notably, many discussions around these values were related to institutional stakeholders rather than publishers. This observation could reflect the already well-established relationships to publishers (in that there is no longer a need to use shared values as a selling point; Section \ref{section_4-2-1}) or the common practice of charging for services in publishing.

\subsubsection{Evolving notions of community and governance} \label{section_4-3-2}
Operating in the community interest is part of Dryad’s value proposition; involving stakeholders in the development of the fee structure was therefore crucial. Community involvement and transparency were leveraged to achieve support from (institutional) stakeholders, demonstrate the legitimacy of the pricing structure by making costs visible, and get further input into customer needs in processes of qualification \parencite{callon_economy_2002}. 
\begin{quote}
    We didn’t go away in a room with just the Dryad team and run the numbers and figure out how much we need to cover our costs. [..] We convened a working group [..] that was mostly comprised of librarian members. We basically opened the books to them and said, look, this is how the organization is structured. Here are the infrastructure costs. You know, curation is a key value. That’s what curation costs. As representatives of the community, what do you want? (P11)
\end{quote}
The composition of these meetings indicates who is seen to be the primary “community” for Dryad. While researchers were originally core members, the community has evolved to focus on stakeholders or customers: actors who will pay for Dryad’s services, including institutions, publishers, and professional societies. To some degree, this evolution is related to the increased size of the Dryad community, and these actors are proxies for researchers, representing researchers’ needs in addition to their own interests. The evolving composition of Dryad’s community is also reflected in governance structures. Dryad’s board includes members with institutional or library expertise and those with publishing backgrounds. Community governance in this case is not about “chasing you down to vote on someone you have never met” (P1), but rather having a selected expert group, developed alongside the business strategy, to steer the organization.
\begin{quote}
    It does make sense when you have a service to have a user group that gives you feedback: is the service evolving in the right direction for it to be useful for the intended purpose? [..] We have this board of directors, which I think historically was a bit of community governance. And now it's kind of developed along the business concept to get people in with different expertise who can help guide and develop Dryad further. (P14)
\end{quote}
The above quotation demonstrates how Dryad’s governance has been shaped by evolutions in financial models. Both the fee structure and the broader business model have also become ways to manage tensions between stakeholders. Previously, different stakeholders were billed using different fee structures, e.g., which caused friction. By providing a consistent and standardized method, the new model is seen as a way to manage both existing and new relationships, becoming a part of the governance framework.
\begin{quote}
    This is finally what I believe is the stable foundation for Dryad to approach growth with confidence because we’re not approaching one institution and saying, yeah, I know they got a different deal, or going to the publishers and saying, I know I can’t explain that. Now we have a much more stable foundation, a more consistent conversation to have, and it creates the ability now to reach into new stakeholders, new markets, new sectors. (P1)
\end{quote}

\section{Discussion and Conclusion: Navigating Tensions} \label{section_5}
We conducted a mixed-methods study of Dryad’s quest for financial sustainability over time. Our findings emphasized that funding open data infrastructure is never a simple process. Rather, it is one which requires reconfigurations of relationships and revenue streams and, for Dryad, processes of assetization. While we have detailed the case of Dryad, sustainably funding open data infrastructures is a system-wide problem that many infrastructures face \parencite{imker_who_2020,skinner_2025_2025}. We therefore discuss our findings in relation to tensions which emerge from our analysis that can be useful for open (data) infrastructures more broadly.

\subsection{Thinking like a Business is Necessary, but It Will Change an Infrastructure.} \label{section_5-1}
“Thinking like a business,” where objects and activities are viewed as potential financial investments, is visible throughout our description of Dryad’s financial evolution, particularly in its recent fee re-structuring (Section \ref{section_4-3}). Dryad and its primary stakeholders (publishers and institutions) engaged in processes of qualification and assetization \parencite{callon_economy_2002,birch_rethinking_2017}, as they co-determined the value of the repository’s offerings through multiple rounds of consultation, feedback, and negotiation. Dryad worked to transform what was identified as a valuable service (curation) into an asset capable of generating durable revenue. This required various activities, such as determining costs and appropriate prices, that were ultimately codified into a new fee structure (Figure \ref{fig:figure_1}). 

Through these activities, Dryad “performed” its business model \parencite{doganova_what_2009}, which required both larger and more minute forms of flexibility \parencite{eschenfelder_financial_2022}, such as “opening the books” to institutional customers, modifying pricing tiers, or moving to a more transactional revenue model (Section \ref{section_4-3}). Implementing the business model in turn shaped core aspects important in sustaining an infrastructure related to value, community and governance. These changes are embedded within Dryad’s new value proposition of being a “service provider operating in the community interest.” Both components of this proposition are interesting to discuss further.

\subsubsection{Dryad’s Value as a Service Provider}
Through the assetization process, Dryad honed articulations of its value to customers as well as its core values. As evidenced by the value proposition, this included strategically focusing on assetizing services, rather than focusing on goods. This reflects the public good characteristics of openness \parencite{david_economic_2003}(Section \ref{section_2-2}), as well as the “rivalrous” nature of services, which do not exhibit these characteristics to the same degree. This move is in line with many prominent recommendations for funding open infrastructure, i.e. charging for services rather than for data access and shifting the burden of payment to data depositors \parencite{organisation_for_economic_co-operation_and_developmentoecd_business_2017,posi_adopters_principles_2025}.

Dryad frames its services differently for different customers. For institutions, Dryad positions itself as a provider of curation services, mobilizing shared community values. For publishers, Dryad provides a service to support meeting data sharing requirements. Our material documenting the assetization of curation services is more detailed than that supporting services for publishers. This could indicate that Dryad did not need to invest in the same level of articulation about its services to publishers, due to their long-standing, established relationships with each other (Section \ref{section_4-2-1}), or that the offering for publishers had yet to be fully developed. While these services were differently framed, they are also not mutually exclusive; organizations such as institutions and publishers can have overlapping values and interests \parencite{malhotra_hybridity_2025}, where institutions also need to support data sharing and publishers also are involved in efforts to curate data. 

\subsubsection{Dryad’s Values: Community Interest and Openness}
The new value proposition also emphasizes operating “in the community interest.” Many of Dryad’s values which we identified in our analysis (Section \ref{section_4-3-1}) resonate with norms of collectivity associated with “the commons” as well as open infrastructures \parencite{moore_care-full_2018,mayer_9_2026}. These include values linked with openness, such as transparency and inclusion, as well as values associated with curation, including human connection and care. These values suggest that “the community” referred to in the value proposition is the broader open infrastructure community, or the scholarly ecosystem itself. Our analysis has shown that assetization led to changes in how Dryad’s community is conceptualized and operationalized in governance structures. While Dryad still aims to connect with individual researchers, its community primarily consists of its stakeholders: institutions and publishers who both (potentially) pay for Dryad’s services and who can guide the development of those services. This was not seen as problematic by our interviewees (and is in line with POSI’s recommendations for stakeholder governance \parencite{posi_adopters_principles_2025}); rather, it was seen to ensure that services, and how they are assetized, evolve in a useful way. It does mean, however, that these primary stakeholders will be the ones to make decisions about and influence Dryad’s future. 

While openness as a normative value emerged as a “unique selling point” for Dryad, particularly in its relation to proprietary infrastructures, it also complicated the development of the fee structure. Openness as a value was seen to be good for philosophical reasons, as Dryad mobilized shared values to forge relationships with institutions and differentiate itself from competitors, but it also made it difficult to convince some customers of the need to pay in a non-collective way. This points to two persistent tensions: first, the tension between the idea that open content and infrastructure should be freely available and the fact that openness requires investment \parencite{skinner_2025_2025}, and second, the friction between transactional and collective funding models for open infrastructure \parencite{moore_care-full_2018}(Section \ref{section_2-2-1}). 

\subsubsection{Transactional vs Collective Models}
Dryad’s new revenue structure is primarily a transactional model including data publication charges (DPCs); the shift to this type of model contributed to the ending of an important relationship in its history (Section \ref{section_4-2-4}). This type of model has parallels in open access publishing, where charging for individual APCs is common. As with APCs, DPCs for individual datasets may be difficult for institutions to budget for, in part because of the dynamics around data production and sharing. The amount of data produced depends on systemic factors, such as third-party funding for research; institutional budgeting timelines may also not match the immediacy of transactional models \parencite{chodacki_not_2026}. Institutions themselves are complex, involving relationships with researchers, librarians, computing staff, and administrative leaders who all have different priorities for managing and funding data sharing. Academic leaders such as provosts, for example, may be more concerned about funding at their own institutions, rather than with the collective benefit which data sharing provides across the scholarly system \parencite{borgman_why_2022}. This could lead to internal friction within an institution about which type of model (collective or transactional) is best for funding data sharing and related infrastructure.

The case of Dryad raises interesting questions around this friction. Can a transactional model have collective benefit and be rooted in open, communal values? The two parts of Dryad’s value proposition suggest that this can be possible. Or, vice versa and perhaps more provocatively, can a collective model also be more transactional in nature, or not serve the collective? Perhaps models implementing aspects of both collectivity and transaction could be a way forward. Recently, \citeauthor{imker_sustainability_2025} has argued that with decreasing institutional funding for research, data repositories might need to balance practicalities with openness, which could mean charging for access in some cases \parencite{imker_sustainability_2025}. As another example, Crossref, a major DOI registration service which charges per DOI, has recently announced that they will waive costs for backlog DOIs to better reflect their goal of creating complete metadata for the scholarly system \parencite{ofiesh_fee_2026}. DataCite, on the other hand, has recently shifted from a transactional model to a collective, membership-fee structure \parencite{buys_new_2026}. This continued flux indicates that these frictions are still very much a matter of debate. 

\subsection{Thinking About Relationships is Part of Thinking like a Business.} \label{section_5-2}
ODIs are often encouraged to cultivate a mix of funding streams \parencite{organisation_for_economic_co-operation_and_developmentoecd_business_2017}; our analysis suggests that having a mix of relationships at various levels of maturity (and timescales) is also important. A combination of established, developing, and “positioning” relationships with customers, competitors and collaborators can help to embed an infrastructure in different webs of relations and practices (Section \ref{section_4-2}). 

Throughout its history, Dryad has carefully reconfigured relationships with other actors by forming such “judicious connections” \parencite{leonelli_philosophy_2023}, rooted in shared values or mutually beneficial goals. Some of these relationships enabled and supported other connections. For example, Dryad’s continued strong relationships with publishers provided the capacities to work on cultivating other relationships and experiment with new funding models, as did shorter-term project funding (Figure \ref{fig:figure_3}). Dryad also carefully positioned itself in relation to other collaborators and competitors: as a replacement for institutional repositories, or, within the GREI initiative, as an important player and peer to other prominent infrastructures (Section \ref{section_4-2-3}). These positioning relationships were perhaps particularly strategic, allowing Dryad to clearly differentiate its place in the scholarly system (e.g. as a generalist rather than specialist) and to define the scope and boundaries of its work. 

Temporalities also emerge as important factors in these considered relationships. Relationships with both publishers and the California Digital Library took significant time to establish; other short-term relationships, within time-defined projects or collaborations, were leveraged to meet specific aims (Section \ref{section_4-2-3}). While it is likely that these (initially) shorter-term relationships extend beyond the lifespan of a project, we also observed relationships which ended or transitioned to new forms (Section \ref{section_4-2-4}). This mirrors broader discussions about when infrastructures themselves should be shut down \parencite{cohn_convivial_2016,posi_adopters_principles_2025}; especially when many similar infrastructures compete for the same funds (Section \ref{section_4-2-3}). While short-term infrastructures could be designed to support particular missions or other longer-term infrastructure \parencite{knopp_temporary_2026}, new relationships, in the form of mergers or consolidations could also be part of the solution for scarce infrastructural funding \parencite{skinner_2025_2025}.

Importantly, our analysis has shown that business models themselves provide ways to manage relationships, perhaps particularly in times of conflict and friction \parencite{doganova_what_2009}(Section \ref{section_4-3}). Through implementing its new business model and process of assetization, Dryad engaged with stakeholders to co-determine the value proposition. Discussions around the fee structure prompted debate, negotiation and compromise, deepening engagement even in frictious situations between different stakeholders with varying interests. Changing from a membership-based model also mediated the ending of Dryad’s extant relationship with the CDL. These examples highlight the importance of conflict and friction within an infrastructure, which, rather than being negative, can be a generative force \parencite{bates_politics_2017}. Engaging around a business model can be a form of dynamic community development, which can lead to (and require) large changes. 

\subsection{Governing Curation Services as an Asset Raises Questions.} \label{section_5-3}
Treating curation services as an asset, something to generate durable income, changes how Dryad relates to and governs this service \parencite{birch_rethinking_2017,birch_assetization_2024}. This decision will impact both how Dryad operates and the broader research data ecosystem. We can speculate about what these changes may be, based on our material. 

For example, assetizing curation services creates a need to make curation costs, which are considerable (Section \ref{section_2-2-3}), more efficient. The focus on cost efficiency will likely require Dryad to be flexible with regards to staffing, which is where Dryad likely has the most agency: There must be enough staff to deliver the service, but curation costs must be covered by revenue. If not enough or too many datasets are received, it could necessitate firing or hiring staff.

The material shows that Dryad also tries to reduce costs of curation services by outsourcing and considering automating parts of it. Dryad is not the only ODI to consider using AI-driven tools, which already have a positive impact on curating collections \parencite{lippincott_sustaining_2026,johnston_understanding_2024} and likely will find wider adoption in the future. Automating and outsourcing services allows Dryad to reduce costs, but this could result in less opportunity to provide human connection and support, which is also seen to be a value of Dryad (Section \ref{section_4-3-1}). Another option could be to explore differentiating “levels of curation” for datasets. Curation activities between and within repositories already vary, due to limited resources and ability to scale curation \parencite{johnston_understanding_2024}. This approach could be integrated into the current business model to keep curation a viable asset, i.e. by providing basic levels of curation for all datasets, while charging more for more extensive levels of curation. This could be a way for Dryad to balance openness with financial sustainability, while still serving public interest \parencite{hook_market_2026}.

The work of curation is often done in service of others and can be seen as a form of care \parencite{davies_care_2024,gabrielsen_gendering_2024}. This raises more philosophical questions: does turning data curation into an asset de-emphasize data curation as care? Caring for data and technologies requires “persistent tinkering in a world full of complex ambivalence and shifting tensions” \parencite[14]{mol_care_2010}. It remains to be seen if data curation as an asset can fit with this idea of “persistent tinkering.” For example, is Dryad's responsibility for a dataset considered fulfilled once the contractually agreed-upon curation measures have been carried out? What would Dryad do if the environment changes and new needs arise for the data under their care? 

\subsection{Concluding thoughts}
The case of Dryad provides an example of one set of practices for working towards financial sustainability and raises questions about the above tensions that can be useful for other ODIs to consider. Our analysis does not suggest that these tensions are negative, or that Dryad should do things differently, but rather highlights the importance of considered relationships and the fact that assetization affects core aspects of an infrastructure. Especially in times of scarce financial resources, ODIs will need to recognize that sometimes large changes are needed to keep an infrastructure operational and decide which changes are acceptable.

\section*{Acknowledgments}
We are grateful to Dryad and to our participants for their support with this study. We would also like to thank our colleagues at the Center for Science and Technology Studies at Leiden University and John Chodacki for their feedback on earlier versions of this article. 

\section*{Disclosure Statement }
Kathleen Gregory’s work on this paper was financed by a Veni grant from the Dutch Research Council (NWO) under the grant number VI.Veni.231S.006.

\section*{Author Contributions}
\textbf{KG}: Conceptualization, Formal analysis, Funding acquisition, Investigation, Methodology, Project administration, Supervision, Visualization, Writing – original draft, Writing – review \& editing
\\
\textbf{DS}: Conceptualization, Data curation, Formal analysis, Investigation, Methodology, Visualization, Writing – original draft, Writing – review \& editing

\theendnotes

\printbibliography

\end{document}